\documentclass[preprint,aps,
showpacs
]{revtex4}

\usepackage{epsfig}

\input{epsf}

\newcommand{\be}{\begin{equation}}
\newcommand{\ee}{\end{equation}}
\newcommand{\bea}{\begin{eqnarray}}
\newcommand{\eea}{\end{eqnarray}}
\newcommand{\nn}{\nonumber \\}
\newcommand{\Bk}{{\bf k}}

\begin{document}

\preprint{Guchi-TP-013}
\date{\today%
}
\title{
Deconstructing Scalar QED at
Zero and Finite Temperature
}

\author{Nahomi Kan}
\email{b1834@sty.cc.yamaguchi-u.ac.jp}
\affiliation{Graduate School of Science and Engineering, Yamaguchi University, 
Yoshida, Yamaguchi-shi, Yamaguchi 753-8512, Japan}

\author{Kenji Sakamoto}
\email{b1795@sty.cc.yamaguchi-u.ac.jp}
\affiliation{Graduate School of Science and Engineering, Yamaguchi University, 
Yoshida, Yamaguchi-shi, Yamaguchi 753-8512, Japan}

\author{Kiyoshi Shiraishi}
\email{shiraish@po.cc.yamaguchi-u.ac.jp}
\affiliation{Graduate School of Science and Engineering, Yamaguchi University, 
Yoshida, Yamaguchi-shi, Yamaguchi 753-8512, Japan}
\affiliation{Faculty of Science, Yamaguchi University,
Yoshida, Yamaguchi-shi, Yamaguchi 753-8512, Japan}

\begin{abstract}

We calculate the effective potential for the WLPNGB in a world with
a circular latticized extra dimension.
The mass of the WLPNGB is calculated from the one-loop quantum effect
of scalar fields
at zero and finite temperature.
We show that a series expansion by the modified Bessel functions is
useful to calculate the one-loop effective potentials.

\end{abstract}



\maketitle


\section{Introduction}

It seems obvious that we lives in the four dimensional world. 
Nevertheless, many theories for unification of forces and/or matter 
in more dimensions than four have been studied\cite{KK}.
A simple possibility is that there is  
a fifth dimension of very tiny size attached to every
point of our four-dimensional world. 
Such an extra dimension can hardly be seen by virtue of its extraordinary
smallness. 

Last year, there appears a novel scheme to describe higher-dimensional
gauge theories,
which is called as deconstruction\cite{ACG,HPW}.
A number of copies of a four-dimensional theory
linked by a new fields can be viewed as a single gauge theory.
The resulting whole theory may be almost equivalent to a 
higher-dimensional theory with discretized extra dimensions.

Recently, Hill and Leibovich pointed out that 
the Wilson line pseudo-Nambu-Goldstone boson (WLPNGB) with low mass
can be naturally obtained by deconstructing five-dimensional
QED\cite{HL1,HL2}. This WLPNGB may be a important candidate for
a cosmological quintessence.

For cosmological application, we should take finite-temperature effect
into account. The behavior of the WLPNGB field may vary
along with the cosmological evolution.

In this paper, we examine the $U(1)$ gauge theory with a
discretized circle. We obtain the effective potential for
the WLPNGB at zero and finite temperature analytically.
Through this paper, we consider the one-loop effect of charged scalar
bosons. Although this model appears unnatural in contrast to the model
with  fermions\cite{HL1,HL2}, the similar technique is valid for the
other models and the application to various models will be studied
elsewhere.

In Sec.~\ref{sec:2}, our model is explained and the mass spectra of the
component fields are shown.
In Sec.~\ref{sec:3}, the one-loop quantum effect of scalar fields is
calculated at zero temperature.
In Sec.~\ref{sec:4}, the one-loop quantum effect of scalar fields is
calculated at finite temperature.
The final section, Sec.~\ref{sec:f}, is devoted to conclusion.

\section{model}
\label{sec:2}

We begin with the lagrangian for deconstructing $(d+1+1)$-D scalar QED:
\bea
{\cal L}&=&\sum_{k=1}^{N}\frac{1}{g^2}
\left[-\frac{1}{4}F_{k}^{\mu\nu}F_{k~\mu\nu}-
(D^{\mu}U_{k})^{\dagger}D_{\mu}U_{k}\right]+
\sum_{k=1}^{N}
\left[-(D^{\mu}\tilde{\phi}_{k})^{\dagger}D_{\mu}\tilde{\phi}_{k}\right]\nn
&+&f\sum_{k=1}^{N}\left(\sqrt{2}\tilde{\phi}_{k}^*U_{k}\tilde{\phi}_{k+1}+
\sqrt{2}\tilde{\phi}_{k}^*U_{k-1}^*\tilde{\phi}_{k-1}-
2f\tilde{\phi}_{k}^*\tilde{\phi}_{k}\right)-
m^2\sum_{k=1}^{N}\tilde{\phi}_{k}^*\tilde{\phi}_{k}\, ,
\eea
where
\be
F_{k}^{\mu\nu}=\partial^{\mu}\tilde{A}_{k}^{\nu}-
\partial^{\nu}\tilde{A}_{k}^{\mu}\, , \qquad
D^{\mu}\tilde{\phi}_{k}=\partial^{\mu}\tilde{\phi}_{k}-
i\tilde{A}_{k}^{\mu}\tilde{\phi}_{k}\, .
\ee
and
\be
D^{\mu}U_{k}=\partial^{\mu}U_{k}-i\tilde{A}_{k}^{\mu}U_{k}+
iU_{k}\tilde{A}_{k+1}^{\mu}\, .
\ee
The label of the fields are considered as periodic modulo $N$,
{\it e.g.}, $\phi_{N+1}\equiv\phi_1$, $\phi_{0}\equiv\phi_N$,
and so on. $N$ is assumed to be larger than $(d+1)/2$.
Usually, the dimension of the {\it space} is taken as three.
We leave the dimensions unfixed because of the possibility
in some combination of compactification schemes in the very early
universe.

We assume that all $U_k$ have a common absolute value $|U_k|=f/\sqrt{2}$.
Hence we can write
\be
U_{k}=\frac{f}{\sqrt{2}}
\exp\left(i\tilde{\chi}_{k}/f\right)\, .
\ee

It is convenient to use the ``Fourier transformed'' modes for the fields:
\be
\tilde{A}_{k}^{\mu}=\frac{1}{\sqrt{N}}\sum_{p}A_{p}^{\mu}
\exp\left[2\pi i\frac{pk}{N}\right]\, ,\qquad
\tilde{\phi}_{k}=\frac{1}{\sqrt{N}}\sum_{p}\phi_{p}
\exp\left[2\pi i\frac{pk}{N}\right]\, ,
\ee
and
\be
\tilde{\chi}_{k}=\frac{1}{\sqrt{N}}\sum_{p}\chi_{p}
\exp\left[2\pi i\frac{pk}{N}\right]\, .
\ee

The fields $A_{p}^{\mu}~(p\ne 0)$ acquire masses by ``absorbing'' the
$\chi_{p}~(p\ne 0)$; the mass spectrum is given by
\be
4f^2\sin^2\left(\frac{\pi p}{N}\right)\, .
\ee
For small $p$, this mass spectrum is approximately given by
\be
f^2\left(\frac{2\pi p}{N}\right)^2\, ,
\ee
which is the Kaluza-Klein spectrum in the continuum theory with
the circle of the circumference $L=N/f$.

The masses of charged bosons are
\be
M_{p}^2=4f^2\sin^2\left(\frac{\pi p}{N}+
\frac{\bar{\chi}}{2f}\right)+m^2\, ,
\ee
where $\bar{\chi}\equiv\chi_{0}/\sqrt{N}$ is a (classically) zero-mode
scalar field.

\section{the effective potential at zero temperature}
\label{sec:3}

\subsection{the one-loop effective potential}

In this section, we compute the quantum effect of the scalar fields
at zero temperature.
The one-loop effective potential for $\bar{\chi}$ is obtained by
\bea
& &\ln\det[-\nabla^2+M^2_p(\bar{\chi})]\nn
&\sim&-\frac{1}{(2\pi)^{d+1}}\sum_{p}
\int_0^{\infty}\frac{dt}{t}~
\int d^{d+1}\Bk~\exp\left[-(\Bk^2+M_{p}^2)t\right]\nn
&=&-\frac{1}{(4\pi)^{(d+1)/2}}\int_0^{\infty}
\frac{dt}{t}t^{-(d+1)/2}~
\sum_{p}\exp\left[-M_{p}^2t\right]\, ,
\eea
after an appropriate regularization.

Using the formula
\bea
\exp\left[-4f^2\sin^2(\theta/2)t\right]&=&e^{-2f^2t}\sum_{\ell=-\infty}^{\infty}
\cos \ell\theta~ I_{\ell}(2f^2t)\nn
&=&e^{-2f^2t}\sum_{\ell=-\infty}^{\infty}
e^{i\ell\theta} I_{\ell}(2f^2t)\, ,
\eea
where $I_{\nu}(x)$ is the modified Bessel function,
we can write the effective potential as
\be
V_{0}(\bar{\chi})=-\frac{2}{(4\pi)^{(d+1)/2}}\sum_{p}\sum_{\ell=1}^{\infty}
\cos {\ell\theta_p}~{\cal I}(\ell;m)\, ,
\ee
where
\be
\theta_p\equiv\frac{2\pi p}{N}+\frac{\bar{\chi}}{f}\, ,
\ee
and
\be
{\cal I}(\ell;m)=\int_0^{\infty}\frac{dt}{t^{(d+3)/2}}
e^{-(2f^2+m^2)t} I_{\ell}(2f^2t)\, .
\label{integ}
\ee
Here the term which is independent of $\bar{\chi}$ is discarded.

Carrying out the summation over $p$ first,
we find that only the term of $p=qN~(q: integer)$ is left.
Then we find
\be
V_{0}(\bar{\chi})=-\frac{2N}{(4\pi)^{(d+1)/2}}\sum_{q=1}^{\infty}
\cos (qN\bar{\chi}/f)~{\cal I}(qN;m)\, .
\ee

\subsection{$m=0$}

First, we examine the case of $m=0$ in detail.
One can find\cite{GR}
\be
{\cal I}(qN;0)=(4f^2)^{\frac{d+1}{2}}
\frac{\Gamma(\frac{d+2}{2})\Gamma(qN-\frac{d+1}{2})}%
{\sqrt{\pi}\Gamma(qN+\frac{d+1}{2}+1)}\, .
\ee
Therefore the effective potential for the WLPNGB is written as
\be
V_{0}(\bar{\chi})=
-\frac{2N\Gamma(\frac{d+2}{2})f^{d+1}}{\pi^{(d+2)/2}}
\sum_{q=1}^{\infty}
\frac{\Gamma(qN-\frac{d+1}{2})}%
{\Gamma(qN+\frac{d+1}{2}+1)}
\cos (qN\bar{\chi}/f)\, .
\ee
In particular, when $d=3$, we obtain
\be
V_{0}(\bar{\chi})=
-\frac{3f^{4}}{2\pi^{2}}
\sum_{q=1}^{\infty}
\frac{\cos (qN\bar{\chi}/f)}%
{q(q^2N^2-1)(q^2N^2-4)}\, .
\ee

Turning back to the case with general $d$, we find that
the effective potential for a large $N$ can be expressed as
\be
V_{0}(\bar{\chi})=
-\frac{2\Gamma(\frac{d+2}{2})}{\pi^{(d+2)/2}L^{d+1}}
\left[\sum_{q=1}^{\infty}
\frac{\cos (qL\bar{\chi})}%
{q^{d+2}}
+\frac{(d+1)(d+2)(d+3)}{24N^2}
\sum_{q=1}^{\infty}
\frac{\cos (qL\bar{\chi})}%
{q^{d+4}}+O(N^{-4})\right]\, ,
\ee
where $L\equiv N/f$.

The mass of the $\chi_0$ field is derived from the effective potential
and turns out to be
\be
m^2_{\chi}=
\frac{2\Gamma(\frac{d+2}{2})\tilde{g}^2}{\pi^{(d+2)/2}L^{d-1}}
Z(d,N)\, ,
\ee
with
\be
Z(d,N)\equiv
\sum_{q=1}^{\infty}
\frac{q^2N^{d+2}\Gamma(qN-\frac{d+1}{2})}%
{\Gamma(qN+\frac{d+1}{2}+1)}
=\zeta(d)+\frac{(d+1)(d+2)(d+3)}{24N^2}
\zeta(d+2)+O(N^{-4})\, ,
\label{ZZ}
\ee
where $\zeta(s)$ is the Riemann zeta function and
$\tilde{g}=g/\sqrt{N}$. Note that the kinetic term of $\chi_0$
includes the factor $\tilde{g}^{-2}$.

\begin{figure}[htb]
\centering
\mbox{\epsfbox{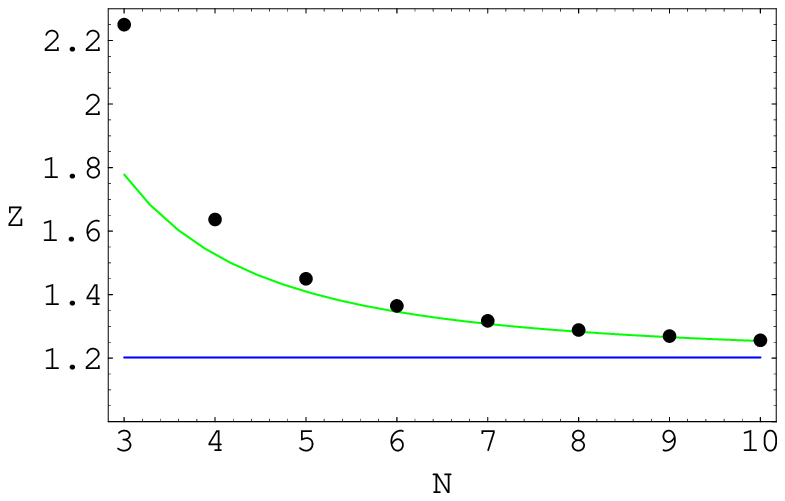}}\\
\mbox{(a)}\\
\mbox{\epsfbox{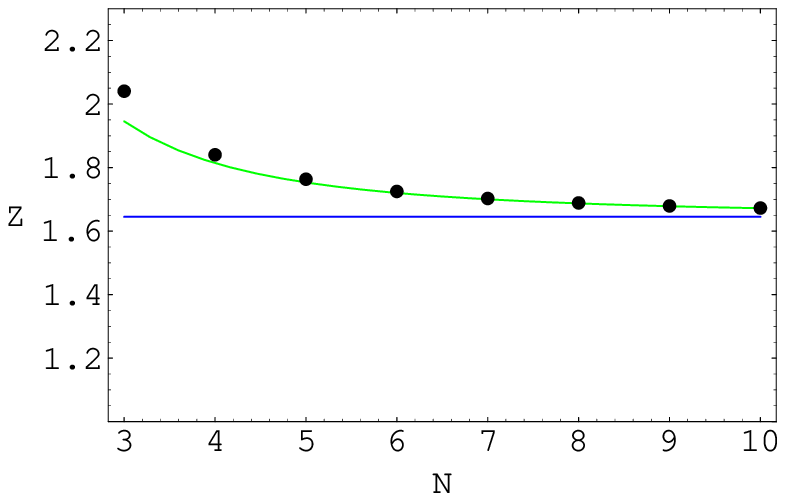}}\\
\mbox{(b)}\\
\bigskip
\caption{%
(a) $Z(3,N)$ is plotted against $N$.
The horizontal line indicates $\zeta(3)$.
The curve illustrates the approximated value up to the order $N^{-2}$.
(b) $Z(2,N)$ is plotted against $N$.
The horizontal line indicates $\zeta(2)$.
The curve illustrates the approximated value up to the order $N^{-2}$.
}
\label{fig1}
\end{figure}

$Z(3,N)$ and $Z(2,N)$ are plotted against $N$ in FIG.~\ref{fig1}.
It is safe to say that the approximation by ``Large $N$ Limit'' is very
good even for $N\approx 10$.

\subsection{$m\gg f$}

For finite $m$, Eq.~(\ref{integ}) reduces to\cite{GR}
\bea
{\cal I}(qN;m)&=&
(2f^2+m^2)^{\frac{d+1}{2}}\left(\frac{f^2}{2f^2+m^2}\right)^{qN}
\frac{\Gamma(qN-\frac{d+1}{2})}{\Gamma(qN+1)}\nn
& &\times {}_2F_1\left(\frac{qN-\frac{d+1}{2}}{2},\frac{qN-\frac{d-1}{2}}{2};
qN+1;\frac{4f^4}{(2f^2+m^2)^2}\right)\nn
&=&
m^{d+1}\left(\frac{f^2}{m^2}\right)^{qN}
\frac{\Gamma(qN-\frac{d+1}{2})}{\Gamma(qN+1)}
{}_2F_1\left(qN-\frac{d+1}{2},qN+\frac{1}{2};
2qN+1;-\frac{4f^2}{m^2}\right)\, ,
\label{mm}
\eea
where ${}_2F_1$ is the Gauss' hypergeometric function.

In the large $N$ limit, ${\cal I}(qN;m)$ behaves as $e^{-mqN/f}$ for large $m$.
For finite $N$, however, ${\cal I}(qN;m)$ approaches to zero not
exponentially but in power law. We find that when 
$m\gg f$ Eq.~(\ref{mm}) reduces to
\be
{\cal I}(qN;m)\sim
m^{d+1}\left(\frac{f^2}{m^2}\right)^{qN}
\frac{\Gamma(qN-\frac{d+1}{2})}{\Gamma(qN+1)}\qquad (m\gg f)\, .
\ee
Thus we obtain
\be
V_{0}(\bar{\chi})\sim-\frac{2m^{d+1}}{(4\pi)^{(d+1)/2}}
\left(\frac{f^2}{m^2}\right)^{N}
\frac{\Gamma(N-\frac{d+1}{2})}{\Gamma(N)}\cos (N\bar{\chi}/f)
\qquad (m\gg f)\, .
\ee
Correspondingly, the mass of the $\chi_0$ field reads
\be
m^2_{\chi}\sim\frac{2\tilde{g}^2m^{d-1}}{(4\pi)^{(d+1)/2}}
\left(\frac{f^2}{m^2}\right)^{N-1}
\frac{N^2\Gamma(N-\frac{d+1}{2})}{\Gamma(N)}
\qquad (m\gg f)\, ,
\ee
which can be very small value if we choose an appropriate value for
$f/m$. This fact suggests that the model can bring about 
interesting models in cosmological application.

\section{the effective potential at finite temperature}
\label{sec:4}

\subsection{the finite-temperature effective potential}

We know that, to study the finite temperature effective potential,
the integration over the frequency is replaced by
the summation over the discrete Matsubara frequencies
(and attach a certain factor)\cite{ft}.
The free energy density is then obtained by
\bea
F&=&-\frac{1}{(2\pi)^{d}\beta}\sum_{p}
\sum_{n'=-\infty}^{\infty}
\int_0^{\infty}\frac{dt}{t}~
\int
d^{d}\Bk~\exp\left\{-\left[\left(\frac{2\pi}{\beta}\right)^2{n'}^2+\Bk^2+
M_{p}^2\right]t\right\}\nn
&=&-\frac{1}{(4\pi)^{(d+1)/2}}\int_0^{\infty}
\frac{dt}{t}t^{-(d+1)/2}~
\sum_{p}\sum_{n=-\infty}^{\infty}
\exp\left[-M_{p}^2t-\frac{\beta^2n^2}{4t}\right]\, ,
\label{jj}
\eea
where $T=\beta^{-1}$ is the temperature.
Obviously, the $n=0$ term in the summation gives the effective potential at zero
temperature.

Now we write $F$ in the form
\be
F=V_0(\bar{\chi})+\Delta V(\bar{\chi})+F_0\, .
\ee
Performing the summation over $p$, one can see that the
finite-temperature  correction to the potential 
$\Delta V(\bar{\chi})$ results in
\be
\Delta V(\bar{\chi})=
-\frac{4N}{(4\pi)^{(d+1)/2}}\sum_{q=1}^{\infty}
\cos\left[qN\frac{\bar{\chi}}{f}\right] ~{\cal T}(qN;m)\, ,
\ee
and
\be
F_0=
-\frac{2N}{(4\pi)^{(d+1)/2}}~{\cal T}(0;m)\, ,
\ee
where
\be
{\cal T}(\ell;m)=\sum_{n=1}^{\infty}\int_0^{\infty}\frac{dt}{t^{(d+3)/2}}~
\exp\left[-(2f^2+m^2)t-\frac{\beta^2n^2}{4t}\right] I_{\ell}(2f^2t)\, .
\ee

Expanding the modified Bessel functions, we can carry out the integration
and obtain
\bea
& &{\cal T}(qN;m)\nn
&=&2(2f^2+m^2)^{(d+1)/2}(f^2/(2f^2+m^2))^{qN}\nn
&\times&\sum_{r=0}^{\infty}
\frac{(f^2/(2f^2+m^2))^{2r}}{r!\Gamma(qN+r+1)}
\sum_{n=1}^{\infty}\left(\frac{\sqrt{2f^2+m^2}\beta
n}{2}\right)^{2r+qN-\frac{d+1}{2}} K_{2r+qN-\frac{d+1}{2}}(\sqrt{2f^2+m^2}\beta n)
\, ,
\eea
where $K_{\nu}(z)$ is the McDonald function
(or the second type of the modified Bessel
function).

\subsection{the high-temperature limit}

In the high-temperature limit
$\beta\rightarrow 0$, the summation over $n$ can be replaced by
integration and the following approximation is obtained:
\bea
{\cal T}(qN;m)&\sim&\frac{\sqrt{\pi}}{\beta}
(2f^2+m^2)^{\frac{d}{2}}\left(\frac{f^2}{2f^2+m^2}\right)^{qN}
\frac{\Gamma(qN-\frac{d}{2})}{\Gamma(qN+1)}\nn
& &\times {}_2F_1\left(\frac{qN-\frac{d}{2}}{2},\frac{qN-\frac{d}{2}+1}{2};
qN+1;\frac{4f^4}{(2f^2+m^2)^2}\right)\nn
&=&\frac{\sqrt{\pi}}{\beta}
m^{d}\left(\frac{f^2}{m^2}\right)^{qN}
\frac{\Gamma(qN-\frac{d}{2})}{\Gamma(qN+1)}
{}_2F_1\left(qN-\frac{d}{2},qN+\frac{1}{2};
2qN+1;-\frac{4f^2}{m^2}\right)\nn
& &\qquad (\beta^{-1}\gg f, m)\, .
\eea
There occurs nothing but the so-called dimensional reduction phenomenon
in high-temperature field theory.

In the case of $m=0$, the high-temperature limit leads to
\be
{\cal T}(qN;0)\sim\frac{1}{\beta}(4f^2)^{\frac{d}{2}}
\frac{\Gamma(\frac{d+1}{2})\Gamma(qN-\frac{d}{2})}%
{\Gamma(qN+\frac{d}{2}+1)}\qquad (\beta^{-1}\gg f)\, ,
\ee
and the effective potential becomes
\be
V(\bar{\chi})\equiv V_0(\bar{\chi})+\Delta V(\bar{\chi})\sim
-\frac{2N\Gamma(\frac{d+1}{2})f^{d}}{\beta\pi^{(d+1)/2}}
\sum_{q=1}^{\infty}
\frac{\Gamma(qN-\frac{d}{2})}%
{\Gamma(qN+\frac{d}{2}+1)}
\cos (qN\bar{\chi}/f)\quad (\beta^{-1}\gg f)\, .
\ee
Particularly, for $d=3$, 
\be
V(\bar{\chi})\sim
-\frac{2f^{3}}{\beta\pi^{2}}
\sum_{q=1}^{\infty}
\frac{N \cos (qN\bar{\chi}/f)}%
{(q^2N^2-1/4)(q^2N^2-9/4)}\qquad (\beta^{-1}\gg f)\, ,
\ee
is obtained.
For general $d$ and large $N$, we find
\bea
V(\bar{\chi})&\sim&
-\frac{2\Gamma(\frac{d+1}{2})}{\beta\pi^{(d+1)/2}L^{d}}
\left[\sum_{q=1}^{\infty}
\frac{\cos (qL\bar{\chi})}%
{q^{d+1}}
+\frac{d(d+1)(d+2)}{24N^2}
\sum_{q=1}^{\infty}
\frac{\cos (qL\bar{\chi})}%
{q^{d+3}}+O(N^{-4})\right]\nn
& &\qquad (\beta^{-1}\gg f)\, ,
\eea
where $L\equiv N/f$.

The mass of the $\chi_0$ field in the high-temperature limit is
\be
m^2_{\chi}\sim
\frac{2\Gamma(\frac{d+1}{2})\tilde{g}^2}{\beta\pi^{(d+1)/2}L^{d-2}}
Z(d-1,N)\qquad (\beta^{-1}\gg f)\, ,
\ee
where $Z(d,N)$ has been defined as Eq.~(\ref{ZZ}).

In the case that $\beta^{-1}\gg m\gg f$, we find
\bea
{\cal T}(qN;m)&\sim&\frac{\sqrt{\pi}}{\beta}
m^{d}\left(\frac{f^2}{m^2}\right)^{qN}
\frac{\Gamma(qN-\frac{d}{2})}{\Gamma(qN+1)}
\qquad (\beta^{-1}\gg m\gg f)\, ,
\eea
and this leads to
\be
V(\bar{\chi})\sim-\frac{2m^{d}}{\beta(4\pi)^{d/2}}
\left(\frac{f^2}{m^2}\right)^{N}
\frac{\Gamma(N-\frac{d}{2})}{\Gamma(N)}\cos (N\bar{\chi}/f)
\qquad (\beta^{-1}\gg m\gg f)\, .
\ee
The mass of the $\chi_0$ field is then
\be
m^2_{\chi}\sim\frac{2\tilde{g}^2m^{d-2}}{\beta(4\pi)^{d/2}}
\left(\frac{f^2}{m^2}\right)^{N-1}
\frac{N^2\Gamma(N-\frac{d}{2})}{\Gamma(N)}
\qquad (\beta^{-1}\gg m\gg f)\, .
\ee
The mass-square of the $\chi_0$ increases with temperature linearly.

\subsection{temperature dependence of the free energy}

In the rest of this section, we investigate
the leading temperature dependence of
the free energy. Though the contribution of the gauge fields
is of course present, we concentrate ourselves only on the 
contribution of the scalar fields.

The dominant dependence on temperature can be found in $F_0$.
Let us remember 
\bea
& &{\cal T}(0;m)=2(2f^2+m^2)^{(d+1)/2}\nn
& &\qquad\times\sum_{r=0}^{\infty}
\frac{(f^2/(2f^2+m^2))^{2r}}{r!\Gamma(r+1)}
\sum_{n=1}^{\infty}\left(\frac{\sqrt{2f^2+m^2}\beta
n}{2}\right)^{2r-\frac{d+1}{2}}
K_{2r-\frac{d+1}{2}}(\sqrt{2f^2+m^2}\beta n)
\, ,
\eea
where we should notice $K_{\nu}(z)=K_{-\nu}(z)$.

At extremely high temperature ($\beta^{-1}\gg f, m$),
the $r=0$ term is dominant and using the limiting form for
a small argument $K_{\nu}(z)\sim
\frac{1}{2}\Gamma(|\nu|)(z/2)^{-|\nu|}$, one obtains
\be
{\cal T}(0;m)\sim
\frac{2^{d+1}\Gamma(\frac{d+1}{2})\zeta(d+1)}{\beta^{d+1}}
\qquad(\beta^{-1}\gg f, m)\, .
\ee
Then this leads to
\be
F\sim F_0\sim
-\frac{2N\Gamma(\frac{d+1}{2})\zeta(d+1)}%
{\pi^{(d+1)/2}\beta^{d+1}}
\qquad(\beta^{-1}\gg f, m)\, .
\ee
This is precisely the free energy for $N$ 
(effectively) massless charged bosons.
This behavior can be derived from the original form of ${\cal T}(0;m)$
\bea
{\cal T}(0;m)&=&\sum_{n=1}^{\infty}\int_0^{\infty}\frac{dt}{t^{(d+3)/2}}~
\exp\left[-(2f^2+m^2)t-\frac{\beta^2n^2}{4t}\right] I_{0}(2f^2t)\nn
&=&\frac{1}{\beta^{d+1}}
\sum_{n=1}^{\infty}\frac{1}{n^{d+1}}
\int_0^{\infty}\frac{dt}{t^{(d+3)/2}}~
\exp\left[-(2f^2+m^2)\beta^2n^2t-\frac{1}{4t}\right]
I_{0}(2f^2\beta^2n^2t)\, ,
\label{oo}
\eea
with the limiting form $I_{0}(z)\sim 1$ for a small argument.

On the other hand, for $\beta^{-1}\ll f$, Eq.~(\ref{oo})
can be approximated, using $I_{0}(z)\sim e^z/\sqrt{2\pi z}$ for a large
argument, as
\be
{\cal T}(0;m)\sim
\frac{2}{\sqrt{4\pi}~f\beta^{d+2}}
\sum_{n=1}^{\infty}\left(\frac{2\beta m}{n}\right)^{\frac{d+2}{2}}
K_{\frac{d+2}{2}}(\beta m n)
\qquad(\beta^{-1}\ll f)\, .
\ee
This leads to
\be
F\sim F_0\sim
-\frac{4N}{(4\pi)^{(d+2)/2}~f\beta^{d+2}}
\sum_{n=1}^{\infty}\left(\frac{2\beta m}{n}\right)^{\frac{d+2}{2}}
K_{\frac{d+2}{2}}(\beta m n)
\qquad(\beta^{-1}\ll f)\, .
\ee
Further, if we assume $m\ll \beta^{-1}$, it is found that
\be
{\cal T}(0;m)\sim
\frac{2^{d+2}\Gamma(\frac{d+2}{2})\zeta(d+2)}%
{\sqrt{4\pi}~f\beta^{d+2}}
\qquad(m\ll \beta^{-1}\ll f)\, .
\ee
Then in this case,
\be
F\sim F_0\sim
-\frac{2\Gamma(\frac{d+2}{2})\zeta(d+2)}%
{\pi^{(d+2)/2}\beta^{d+2}}\frac{N}{f}
\qquad(m\ll\beta^{-1}\ll f)\, ,
\ee
is obtained.
This result coincides with the one of the finite-temperature
continuum Kaluza-Klein theory with circle length $L=N/f$\cite{RR},
after replacing the scalar degree of freedom.

We have found that $(-F)$ behaves as $T^{d+1}$ at high temperature,
while it behaves as $T^{d+2}$ at lower temperature than $f$.
This fact indicates that the dimension of the spacetime seems $d+2$
for $T<f$ and again $d+1$ for $T>f$.
Of course, at very low temperature $T\ll f/N$,
as one can see from Eq.~(\ref{jj}) for $m=0$ and $\bar{\chi}=0$,
\be
F\sim
-\frac{2\Gamma(\frac{d+1}{2})\zeta(d+1)}%
{\pi^{(d+1)/2}\beta^{d+1}}
\qquad(\beta^{-1}\ll f/N)\, .
\ee
So, we recognize the world as $(d+1)$-dimensional spacetime
with the lowest-mode field at very low temperature.

\section{conclusion}
\label{sec:f}

In conclusion, the effective potential for the WLPNGB 
in scalar QED with a discretized dimension has been calculated
at zero and finite temperature.
We have utilized the expansion in terms of the modified Bessel functions,
which is also useful for computing the one-loop effect in models
with more discretized (or, latticized) dimensions\cite{fw}.
We have found that approximating the one-loop effect by large $N$
expansion is valid if the model has a limiting form of infinite $N$.

We have also found that the $(mass)^2$ of the WLPNGB increases linearly
with high temperature. This serves some possibilities: 
Coherent oscillations of the WLPNGB field
may change the frequency according to the expansion of the universe,
or,
if domain walls might be produced, their mass density decreases
as temperature decreases.
Furthermore, novel temperature-dependence of energy density may
bring interesting consequences to the very early universe.
These cosmological implications will be clarified
after analyzing more realistic models and incorporating other
matter fields.

We should consider the one-loop effect of fermions
for more natural particle theory.
Moreover degenerate fermions may largely affect the WLPNGB mass
and the entire potential.
These subjects will be discussed elsewhere\cite{mi}.


\begin{acknowledgments}
We would like to thank Y. Cho for his valuable comments
and for the reading the manuscript.
\end{acknowledgments}



\end{document}